\documentstyle[prl,twocolumn,aps]{revtex} 
\input{epsf} 
 
\begin{document} 
\draft 
\twocolumn 
\title{Semiclassical theory of the emission properties of 
wave-chaotic resonant cavities} 
 
\author{Evgenii E. Narimanov,$^{1}$ Gregor Hackenbroich,$^{2}$ 
Philippe Jacquod,$^{3}$ and 
A.~Douglas Stone$^{3}$ } 
 
\address{$^{1}$ Bell Laboratories-- Lucent Technologies, 
700 Mountain Ave., Murray Hill NJ 07974} 
\address{$^{2}$ Universit\"at GH Essen, Fachbereich 7, 45117 Essen, 
Germany} 
\address{$^{3}$ Department of Applied Physics, P.O.Box 208284, 
Yale University, New Haven, CT 06520-8284} 
 
\date{ \today} 
 
\maketitle 
 
\begin{abstract} 
We develop a perturbation theory for the lifetime and emission 
intensity for isolated resonances in asymmetric resonant cavities. 
The inverse lifetime $\Gamma$ and the emission intensity $I(\theta )$ in 
the open system are expressed in terms of matrix elements of 
operators evaluated with eigenmodes of the closed resonator. These 
matrix elements are calculated in a semiclassical approximation 
which allows us to represent $\Gamma$ and $I(\theta )$ as sums 
over the contributions of rays which escape the resonator by 
refraction. 
\end{abstract} 
\vspace*{-0.05 truein} 
\vspace*{-0.15 truein} 
 
The semiclassical theory of wave equations with non-integrable 
classical (short wavelength) limits has been a topic of great 
interest over the past decade. Much progress has been made for closed 
systems based on periodic orbit theory, particularly in the case of 
fully chaotic systems with no stable periodic orbits or families of 
regular orbits (KAM tori) \cite{Gutzwiller}.  One may for example obtain 
eigenvalues 
with an energy resolution inversely proportional to the length of 
the orbits summed \cite{Smilansky}.  Much current work addresses the 
generic case of quantum/wave systems with mixed dynamics in the 
classical limit, i.e. with stable, unstable and regular regions in 
phase space \cite{Reichl}.  Here we focus on the situation where a system with 
mixed dynamics is coupled to an infinite asymptotic region and 
the eigenstates become resonances with complex eigenvalues 
\cite{Gaspard,Shepelyansky}.  A crucial quantity to calculate in this case 
is the resonance width (imaginary part of the eigenvalue). 
A basic problem here for semiclassical approaches is that the 
resonance width may be exponentially smaller than the level spacing, making 
a direct periodic orbit method very difficult. 
 
Although our approach could be applied to quantum hamiltonians, we will 
focus here on an important application for such a theory: the formally 
similar problem of 
dielectric microcavities. Such resonators are used as high-Q resonators for 
microlasers and optical spectroscopy \cite{Slusher,Ilchenko}. If such 
resonators are rotationally symmetric the wave equation is separable 
and analytic solutions for resonance properties are possible; 
moreover the corresponding ray dynamics (within the resonator) is 
integrable with conserved 
angular momentum. When such resonators are deformed from rotational 
symmetry the ray dynamics is generically mixed and no analytic 
results were known for resonance properties, except for very small 
deformations which can then be treated perturbatively 
\cite{Youngpert}.  However some time ago \cite{OL1,PRL,OL2,Nature} it was 
noted that 
such asymmetric resonant cavities (ARCs) are equivalent to quantum 
billiards with refractive escape and a ray-optics model was 
developed for the lifetime and emission pattern. Although quite 
successful for low-index resonators where only whispering gallery 
modes are important, this ray model has several limitations.  1) It 
only describes quantitatively the whispering gallery modes of such 
resonators.  2) It neglects all interference effects. 3) It is valid 
only near a certain adiabatic limit for the phase space flow 
\cite{OL2,Nature}. 
However recent experimental results on deformed cylindrical 
semiconductor microlasers (a high-index, high-gain system) 
demonstrated high-power and highly directional emission from 
"bow-tie" modes which are {\it not} well described by this 
model\cite{science}.  Therefore there is both general theoretical and 
specific experimental motivation for the semiclassical theory of ARCs which 
we present below. 
 
The key simplification we employ to make the problem tractable is to 
assume that the eigenstates of the "closed" system are known and 
develop a theory which is perturbative in the coupling to the 
outside {\it but not in the deformation}.  The resulting analytic 
formulae are convenient to use in combination with either numerical 
or semiclassical calculations for the closed problem, and have a 
simple physical interpretation in terms of refractively escaping 
rays.  For the case of modes associated with {\it stable} orbits of 
the closed system a full analytic semiclassical solution can be 
found which depends only on $nkR_0$ ($k$ is the wavenumber of the 
radiation, $R_0$ is the average radius of the 
ARC and n is the index of refraction) and properties of the classical orbit 
(e.g. its period and stability).  Finally we note that the numerical 
solution of such problems becomes intractable due to either instabilities 
or convergence problems  \cite{Barton} when $kR_0$ is of order 100, 
well below the range of many optical experiments.  In this case 
the semiclassical theory presented here is (to our knowledge) the 
only available method of solution. 
 
We consider a long cylindrical dielectric resonator of uniform index n with 
a non--circular cross--section parametrized by the 
distance $R(\phi)$ between the origin and a point on the the 
boundary in the {$\phi$}--direction. We consider the TM polarization 
(assuming $k_z=0$) 
for which the electric field is perpendicular to the cylinder axis and 
both the field and its derivative are continuous at the boundary. A 
quasibound state with complex frequency $\omega = n c k$ can be written 
in the form: 
\begin{eqnarray} 
E\left({\bf r}, t\right) = 
e^{- i c k t} 
\sum_{m=-\infty}^\infty i^m  A_m\left(kr\right) e^{ i m \phi} , 
\label{eq:psi_general} 
\end{eqnarray} 
where 
\begin{eqnarray} 
 A_m & = & 
\left\{ 
\begin{array}{ll} 
 \alpha_m  H_m^{+}\left(n k r\right) 
+  \beta_m H_m^{-}\left(n k r\right)  & {\rm if}\  r \leq R(\phi) ,\\ 
\gamma_m H_m^{+}\left(k r\right)  & {\rm otherwise}. 
\end{array} 
\right. 
\nonumber 
\end{eqnarray} 
where $H_m^{\pm}$ are Hankel functions. 
Such resonant solutions can only exist for complex $k$ as there is 
no incoming wave for $r > R(\phi)$; the imaginary part $\Gamma = - 
{\rm Im} k$ is the resonance width. 
 
Defining coefficient vectors $\left| \alpha \rangle \right. $, $ 
\left| \beta \rangle \right. $, $ \left| \gamma \rangle \right. $ 
from Eq.~(\ref{eq:psi_general}), the regularity of the wavefunction at 
$r = 0$ requires 
$\left| \alpha \rangle \right.  = \left| \beta\rangle \right. $. 
Using the continuity conditions to eliminate $\left| \gamma \rangle 
\right.$ and $\left| \beta \rangle \right.$ the resonance condition can 
be expressed as the non--hermitian eigenvalue problem 
\begin{eqnarray} 
\left[ {{\cal H}_0}\left(k\right) + 
{V}\left(k\right) \right] 
\left| \alpha \rangle \right. = 0 , 
\label{eq:alpha_general} 
\end{eqnarray} 
where ${\cal H}_0$ and $V$ are the hermitian and antihermitian part, 
respectively, of the matrix ${\cal M} \equiv \left(J' J'\right) 
-(1/n) ( J' {H^{+}}') \left( H^- H^+ \right)^{-1} \left( H^- J 
\right)$. Here, 
${\cal M}$ is defined in terms of matrices $(Z 
\bar{Z})$ with matrix elements 
\begin{eqnarray} \left( Z \bar{Z} \right)_{m \ell } = 
\frac{1}{2 \pi} 
\int d\phi \ 
Z_m\left(k_0\right) 
\bar{Z}_{\ell} \left( k  \right) 
\ 
e^{ i \left(\ell - m\right) \phi } , 
\label{eq:me} 
\end{eqnarray} 
where $Z_m(k)$ and $\bar{Z}_m(k)$ stand for either 
$H^{\pm}_m\left(k R\left(\phi\right)\right)$, the Bessel function $J_m \left(n k R\left(\phi\right)\right)$, or their derivative, and $k = k_0 - i 
\Gamma$. 
 
The perturbation theory is based on the observation that for  a 
narrow resonance $\Gamma \ll \Delta$, (where $\Delta $ 
is the resonance spacing), the antihermitian term $V$ is a small correction 
in Eq.~(\ref{eq:alpha_general}) which may be expanded in powers of 
$\Gamma$, to obtain to leading order in $V$: 
\begin{eqnarray} 
\Gamma & = & { \langle \alpha^{(0)} \left| V \right| \alpha^{(0)} 
\rangle \over  \langle \alpha^{(0)} \left| 
\left({\partial {\cal H}_0}/{\partial k} \right) 
\right| 
\alpha^{(0)} \rangle } ,
\label{eq:gamma_pert} 
\end{eqnarray} 
where the matrix elements are taken at the real wave number 
$\bar{k}_0$ defined by the hermitian eigenvalue problem 
\begin{eqnarray} 
{\cal H}_0\left(\bar{k}_0\right) | \alpha^{(0)} \rangle  = 0 
. \label{eq:closed} 
\end{eqnarray} 
The operator ${\cal H}_0$ describes a closed resonator with boundary 
conditions to be specified below. Note, that the 
denominator in Eq.~(\ref{eq:gamma_pert}) depends only on the 
properties of the closed system, and can be regarded as a 
normalization factor. 
 
Similarly to (\ref{eq:gamma_pert}), one can express the far-field 
emission intensity from the resonance in terms of the eigenstates of 
${\cal H}_0$, 
\begin{eqnarray} 
I\left(\theta\right) = | \sum_{m, \ell, \ell'} 
e^{ i \ell' \theta } 
\left(H^- H^+\right)^{-1}_{m \ell} 
\left(H^- J\right)_{\ell \ell'} 
\alpha^{(0)}_m |^2 . 
\label{eq:intensity_pert} 
\end{eqnarray} 
 
These results represent a general perturbation theory for the 
problem requiring only small $\Gamma$ and may be of value in their current 
form.  However, far more interesting results can be obtained 
by taking advantage of the semiclassical limit in which the 
wavelength of the radiation is assumed much smaller than the average 
radius of the resonator $nkR_0 \gg 1$.  This condition is satisfied in 
essentially all optical experiments. In this limit Hankel and Bessel 
functions may be well-represented by the "approximation by tangents" 
\cite{Abramowitz} 
and the matrix elements and angular momentum sums in 
(\ref{eq:gamma_pert}), (\ref{eq:intensity_pert}) can be evaluated in 
the stationary phase approximation. In this approximation, the 
boundary condition corresponding to (\ref {eq:closed}) becomes 
 
\begin{eqnarray} 
\left[\nabla_n  - i \sum_m k_n 
\exp\left(i \varphi \right) 
\left| m \rangle \langle m \right|\right] \left| E \rangle 
\right. = 0 , 
\label{eq:bc_closed} 
\end{eqnarray} 
where $| m \rangle$ is an eigenfunction of angular momentum about the 
cylinder axis, $\nabla_n$ is the normal derivative at the boundary of the 
resonator and $k_n(m,\phi)$ the normal component of the wavevector 
$\bf k$, and $\varphi\left(k_n\right)$ the Fresnel phase change upon 
reflection from a flat dielectric boundary.  When $n \gg 1$, Eq. 
(\ref{eq:bc_closed}) reduces simply to von Neumann boundary conditions. 
 
We first present the results for $I(\theta)$. 
Evaluating Eq. (\ref {eq:intensity_pert}) in the stationary phase 
approximation gives three conditions: 1)  The matrix $(H^-H^+)^{-1}$ 
is diagonal.  2) 
An internal angular momentum $m$ only couples to the angular momentum $l$ which 
allows the corresponding ray to satisfy Snell's law (either in transmission 
or reflection). 3) This ray with angular momentum $\ell_0$ must be 
emitted into 
angle $\theta$ in the far-field (see Fig.~1a).  The last two 
conditions imply that there exists at most one such ray with angular 
momentum $\ell_0$ for each value of $m,\theta,kR$. If $m$ corresponds to 
an angle of incidence $\sin \chi \approx m/nkR$ which is totally-internally 
reflected $(m/kR > 1)$ then there is no such ray, and the 
corresponding component of the 
field $\alpha_m$ does not contribute to emission, hence $I(\theta)$ 
is only due to refracted rays.  At large index the contribution from 
evanescent escape (tunneling) will be important and is not described by 
this level of approximation. One finds 
for the (non--normalized) emission intensity 
\begin{eqnarray} 
I\left(\theta\right) & = & \left| \ \Sigma_m \exp(im\phi_{\pm}) 
\ e_m\left(\theta\right) 
\alpha_m^{(0)}\ 
\right|^2, 
\label{eq:intensity_sc} 
\end{eqnarray} 
where $e_m\left(\theta\right) = \sum_{\pm} B\left(\phi_\pm \right) 
\exp ( i \ell_0 \left(\theta - \phi_\pm - \pi /2 \right) )$, and 
\begin{eqnarray} 
B\left(\phi_\pm \right) = 
\frac{H_{\ell_0}^-\left(kR\right) H_m^\pm\left(nkR\right)} 
     {H_{\ell_0}^-\left(kR_0\right) H_{\ell_0}^+\left(kR_0\right)} 
\left[ 1 - 2 \frac{R'}{R} \tan\left(\theta - \phi_\pm \right) 
\right.\ \ \ \ \ \nonumber \\ 
 +  \left. \frac{\sec\left(\theta - \phi_\pm \right)}{kR} 
\left[ \frac{{R''}}{{R'}} \left(\ell_0 - m\right) 
\pm 
\frac{ \left(R'/R\right)^2 m^2}{\sqrt{\left(nkR\right)^2 - m^2}} 
     \right] \right]^{-1/2} . 
\nonumber 
\end{eqnarray} 
Here, $\ell_0 = k R \sin\left(\theta - \phi_\pm \right)$ where the 
$\phi_\pm$ are the two distinct points on the boundary where Snell's law is 
satisfied for the incident and reflected ray (see Fig.~1a, only incident 
ray is shown). 
 
Thus Eq.~(\ref{eq:intensity_sc}) has a simple physical 
interpretation: $\{\alpha_m \}$ describe the relative weight for each 
angular momentum component in the closed resonator, while the 
$e_m$ are refraction amplitudes describing the probability amplitude 
for refractive escape from angular momentum $m$ in the direction 
$\theta$.  If the interference terms are neglected in Eq. 
(\ref{eq:intensity_sc}) the result is essentially the ray model of 
references \cite{OL1,PRL,OL2,Nature} generalized for an arbitrary initial 
state, and provides a more rigorous justification for that 
model.  However if one maintains (as we do in Fig.~1) the 
generality of Eq.~(\ref{eq:intensity_sc}) including the cross-terms 
then the interference effects neglected in the ray model are 
captured. 
 
\begin{figure}[t] 
\begin{center} 
\leavevmode 
\epsfxsize = 8.cm 
\epsfbox{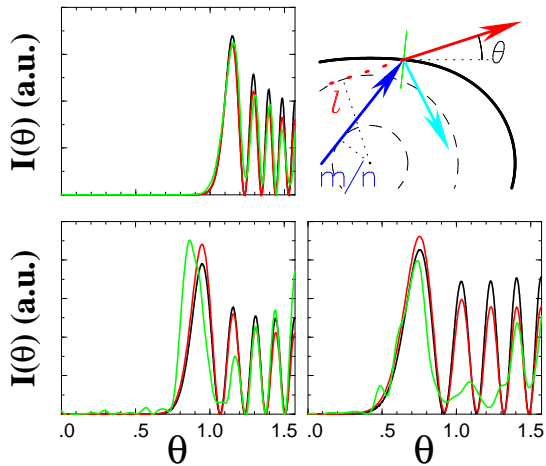} 
\end{center} 
\vspace*{-4.5cm}
\caption{Counterclockwise from top right: (a) Schematic of the
classical ray emission. (b-d)
Far-field emission intensity for a low angular momentum (``bouncing ball'')
mode with
$nkR \approx 150.$,  refractive index $n=1.5$ (b), 2.5 (c), 3.3 (d)
and  deformation of $\epsilon=0.106$.
The black, red and green curves show the exact, perturbative
(\protect\ref{eq:intensity_pert})
and semiclassical (stationary phase) (\protect\ref{eq:intensity_sc})
emission intensities respectively.
}
\end{figure} 
 
In Fig.~1 we show a test of the emission directionality of 
Eq.~(\ref{eq:intensity_sc}), comparing  to an exact numerically-generated 
resonance in 
the resonator with $10\%$ ($\epsilon \approx 0.1$) quadrupolar 
deformation $R(\phi) =R_0 ( 1 + \epsilon \cos(2\phi))$, for 
different values of the refractive index.  The exact result (black curve) is 
very well reproduced by the 
simple perturbative result of Eq.~(\ref{eq:intensity_pert}) 
(red curve) over the whole range of indices of refraction, whereas the 
semiclassical approximation (Eq.~(\ref{eq:intensity_sc}), the green curve) 
works 
well for the lower indices but not so well for larger index.  The origin of 
this trend is the neglect of the "non-classical" escape 
processes mentioned above, which become more important at 
high index.  These tunneling processes are similar to the "ghost" 
contributions in periodic orbit theory \cite{ghosts} and can be included by 
going 
beyond the real saddles in the stationary-phase approximation as is done in 
that context.  We defer this refinement to future work. 
 
Our theory reduces the calculation of $\Gamma$ and $I(\theta)$ to 
solving for the eigenstates of the appropriate closed resonator and 
substituting into Eqs.~(\ref{eq:gamma_pert}) and 
(\ref{eq:intensity_pert}).  In many cases it is possible either to 
evaluate those eigenstates analytically (e.g.\ for island or torus 
states) or to model their statistical properties and hence obtain 
information about lifetimes and emission patterns in chaotic 
resonators. As an example, we shall consider the eigenstates 
localized at the stable islands of the mixed phase space 
of an ARC. Such eigenstates are of key importance for the recently 
studied deformed semiconductor microlasers\cite{science}, where a 
factor of $10^3$ enhancement of lasing power was attributed to the 
emission from a mode localized at the stable bow-tie periodic 
orbit (see inset in Fig.~2a). 
 
We obtain the wave function at the boundary of the resonator using the 
semiclassical theory developed in Ref.~\cite{physicad}. The 
quantization on stable islands yields harmonic oscillator wave 
functions; the corresponding oscillator frequency is determined by 
the classical dynamics in the vicinity of the periodic orbits. For 
the ground state resonance, localized at the center of a chain of 
stable island, we obtain the width 
\begin{eqnarray} 
\Gamma & = & \frac{R_0}{2 \sqrt{\pi} n^2 } \sum_\mu 
{b \over R_\mu \cos^2\chi_\mu} \int_{-1}^1 dx 
e^{- i {b}^2 
\left( x - n \sin\chi_\mu\right)^2 }, 
\label{eq:gamma_island} 
\end{eqnarray} 
where $\chi_\mu$ is the periodic orbit incidence angle and $R_\mu$ 
the distance from the bounce point $\mu$. The parameter $b$ measures 
the "spread" of the wave function in angular momentum. Choosing as 
phase space coordinates the arc lenth along the boundary and the 
transverse momentum $\sin \chi$, one finds $b = (k a_\mu)/ \sqrt{1 + 
  a_\mu^4 \left(\eta - f\right)^2}$, where $ f = 
 s_\mu R_{\mu}''/R_{\mu} 
  + \left(R_{\mu}'/R_{\mu}\right)^2 \ell^2 / s_\mu$, 
the angular momentum $\ell$ corresponds to the refracted 
ray, related to the periodic orbit $\mu$ by $\sin\chi_{\ell} = n 
\sin\chi_\mu$, and
$s_\mu \equiv \sqrt{\left(k R_\mu\right)^2 - \ell^2 }$.
The parameters $a_\mu = 
\left(4 - {\rm Tr}\left[ m_{ij}\right]\right)^{1/4} \left(2 
  m_{21}^\mu \right)^{-1/2}$ and $\eta_\mu = \left(m_{22}^\mu - 
  m_{11}^\mu\right)/ m_{21}^\mu$ are  defined
 in  term 
of  the  monodromy matrix \cite{Reichl} $m_{ij}^\mu$.
 We note that our semiclassical method is justified as 
long as the area in the phase space ``covered'' by the eigenstate, 
is significantly smaller than the size of the islands. 
 
Similar to the calculation of $\Gamma$, we derived a closed 
semiclassical expression for the emission intensity $I(\theta)$.  This 
calculation will be presented elsewhere \cite{unpub}, one result for the 
case of the bow-tie resonance is shown in Fig. 2b. 
 
The comparison of the semiclassical emission intensity and width with 
the exact calculation is presented in Fig.~2. Note, that, as 
indicated by the Husimi projection of the exact eigenstate (panel 
(a)), the phase space area of the island is of the order of the 
effective $\hbar = 1/nkR_0$. The resulting leakage of amplitude 
from the island should therefore lead to deviations of the actual 
intensity from the semiclassical prediction. However, the general 
structure of the far--field emission pattern (panel (b)) is well 
reproduced by the semiclassical result.

\begin{figure}[t] 
\begin{center} 
\leavevmode 
\vspace*{-1.1cm} 
\epsfxsize = 16.3cm 
\epsfbox{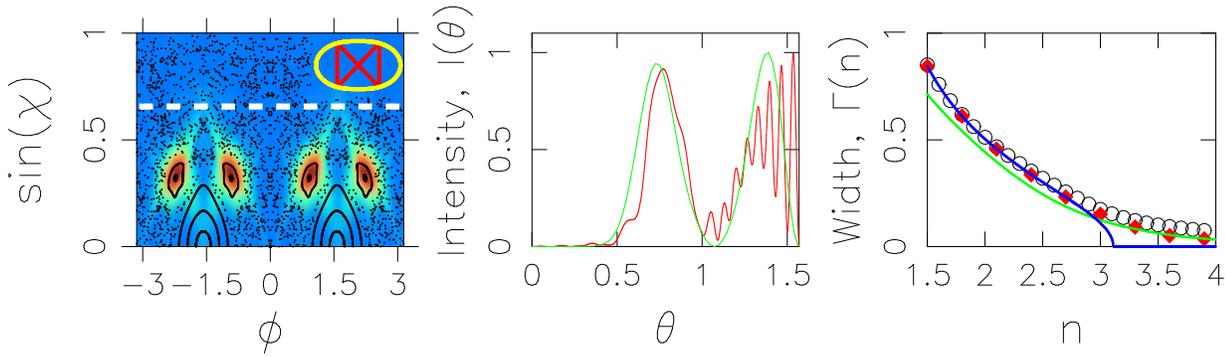} 
\end{center} 
\caption{The bow-tie resonance, its emission pattern, and its width. 
(a) Husimi projection (color scale - red denotes high intensity) of bow-tie 
mode at $n = 1.5$, $n k R = 102.$, for quadrupolar 
defomation $\epsilon = 0.168$ in $(\phi, \sin\chi)$ coordinates, superimposed 
on the corresponding Poincar\'e Surface of Section. 
The white dashed line indicates the threshold for classical emission, 
$\sin\chi_c=1/n$, 
above which only tunneling escape occurs. 
Inset: Schematic of bow-tie periodic orbit. (b) Far-field emission: Exact 
numerical results (red), and the analytic 
semiclassical result for this island state (green).  The formula leading to 
the green curve is not given in the text due to space limitations, it will 
be derived elsewhere {\protect \cite{unpub}}. 
(c) Comparison of the exact width of the bow-tie resonance 
(circles and diamonds) with the semiclassical (green line) 
expression (\protect\ref{eq:gamma_island}) and classical (blue line) 
prediction.  Black circles and red diamond correspond to bow-tie resonances 
close to $nkR_0=45.$ and $nkR_0=102.$ respectively.} 
\end{figure}

{\samepage In panel (c) we compare the results for the resonance width. In 
addition to the exact data points (circles and diamonds) and the 
semiclassical prediction of Eq.~(\ref{eq:gamma_island}) (red curve), 
we also show the classical result $\Gamma = (2 R_0/L) \ln r$ (blue 
curve), based on the Fresnel reflection coefficients $r$, calculated 
at the incidence angle of the periodic orbit ($L$ 
here is the total length of the periodic orbit). The simple classical model 
works very well at index near unity since it is not perturbative in 
$\Gamma/\Delta$ which is not very small in this case; whereas the 
semiclassical 
theory of 
Eq. (\ref{eq:gamma_island}) is perturbative and shows a small but visible 
discrepancy in this range.  However once the index becomes large enough 
that a ray on the bow-tie orbit will be totally-internally reflected the 
classical model gives the unphysical result $\Gamma =0$, whereas our 
semiclassical model continues to decrease smoothly and in good agreement 
with the numerical data.}

\pagebreak 
 
\vspace*{4.0cm} 
 
In summary, we have developed a theory of resonance lifetime and 
emission intensity in nonintegrable dielectric resonators. The 
theory is in a good agreement with numerical data, has a simple 
physical interpretation in terms of refractive emission, and gives 
non-trivial predictions for the lifetimes and emission patterns in 
asymmetric resonant cavities. 
 
We gratefully acknowledge the support of the NSF grant PHY9612200, the Deutsche Forschungsgemeinschaft, the Swiss National Science Foundation, 
and the Aspen Center for Physics.  We thank Jens Noeckel for useful 
conversations.

\end{document}